\documentstyle[aps,multicol,epsf]{revtex}
\newcommand{\mib}[1]{\mbox{\boldmath $#1$}}

\newcommand{\Ohm}{\Omega}
\newcommand{\cm}{{\rm cm}}
\newcommand{\citeScience}[3]{Science\ {\bf #1}, #3 (#2)}
\newcommand{\citePR}[3]{Phys.\ Rev.\ {\bf #1}, #3 (#2)}
\newcommand{\citePRB}[3]{Phys.\ Rev.\ B\ {\bf #1}, #3 (#2)}
\newcommand{\citePRL}[3]{Phys.\ Rev.\ Lett.\ {\bf #1}, #3 (#2)}
\newcommand{\citeJPSJ}[3]{J.\ Phys.\ Soc.\ Jpn.\ {\bf #1}, #3 (#2)}
\newcommand{\citePTP}[3]{Prog.\ Theor.\ Phys.\ {\bf #1}, #3 (#2)}

\begin{document}
\draft

\title{Spin Chirality Fluctuation and Anomalous Hall Effect in Itinerant 
Ferromagnets}
\author{Shigeki Onoda}
\address{Tokura Spin Superstructure Project, ERATO, Japan Science and 
Technology Corporation, Department of Applied Physics, University of Tokyo, 
Tokyo 113-8656, Japan}
\author{Naoto Nagaosa}
\address{Department of Applied Physics, University of Tokyo, Tokyo 113-8656, 
Japan}
\address{Correlated Electron Research Center, AIST, Tsukuba, Ibaraki 305-0046, 
Japan}
\date{Recieved}
\maketitle
\begin{abstract}
The anomalous Hall effect due to the spin chirality order and fluctuation is studied theoretically in a Kondo lattice model without the relativistic 
spin-orbit interaction. Even without the correlations of the localized spins, $\sigma_{xy}$ can emerge depending on the lattice structure and the spin anisotropy. We reveal the condition for this chirality-fluctuation driven mechanism for 
$\sigma_{xy}$. Our semiquantitative estimates for a pyrochlore oxide Nd$_2$Mo$_2$O$_7$ give a finite $\sigma_{xy} \sim 10\ \Ohm^{-1} \cm^{-1}$ together with a high resistivity  $\rho_{xx} \sim 10^{-4}-10^{-3}\ \Ohm\ \cm$, in agreement with experiments. 
\end{abstract}
\begin{multicols}{2}
\narrowtext
The quantum charge transport in the background of the (fluctuating) spins is one of the central issues in strongly correlated electronic systems~\cite{AndersonHasegawa55,BaskaranAnderson88,Laughlin88,WenWilczekZee89,LeeNagaosa92,dagotto}. This issue is relevant to most of the intriguing systems such as high-$T_{\rm c}$ cuprates~\cite{BednorzMuller86}, manganites showing colossal magnetoresistance (CMR)~\cite{CMR_review}, and magnetic semiconductors. Of particular interest is the quantum Berry phase generated by the solid angle subtended by the spins in the noncoplanar configurations, which is called spin chirality~\cite{BaskaranAnderson88,Laughlin88,WenWilczekZee89,LeeNagaosa92}. Recently it has been proposed that this spin chirality contributes to the anomalous Hall effect (AHE) in ferromagnet because the spin chirality acts as an effective magnetic field or gauge
flux ${\vec b}$ for the charge carriers~\cite{ong,YeKimMillisShraimanMajumdarTesanovic99,Lyanda-Geller01,OhgushiMurakamiNagaosa00}. These contrast with the conventional theories of the AHE, where the spin-orbit coupling is indispensable. Conventional theories can be roughly categorized into two. One is based on the band picture where the interband matrix elements between the Bloch wave states with the spin-orbit coupling gives the finite Hall conductivity $\sigma_{xy}$~\cite{KarplusLuttinger54}. This theory has recently been given modern interpretations in terms of Berry phase, parity anomaly and Chern numbers~\cite{monoda}. The other is based on the skew scattering of itinerant electrons by fluctuating localized moments~\cite{Kondo62}.

Experimentally the AHE in Nd${}_2$Mo${}_2$O${}_7$ has been discussed from the viewpoint of the spin chirality~\cite{TaguchiScience01,YasuiSato}. Among other Mo pyrochlore systems, this compound exhibits a noncoplanar spin structure and a large $\sigma_{xy}$ in the metallic ferromagnetic state below the Curie temperature $T_{\rm C}\sim90$ K. $\sigma_{xy}$ at a lowest temperature has been explained consistently in terms of a mean-field theory by taking into account the slight tilting of Mo spins coupled to the Nd spins~\cite{TaguchiScience01}. However, the temperature dependence of $\sigma_{xy}$ is much slower than that of the transverse ordered component $M_{\perp}$ of the Nd spins, which can not be explained in the mean field theory~\cite{TaguchiScience01}. This experimental fact, together with the spin-ice nature of the Nd spins, strongly suggests that $\sigma_{xy}$ is generated by the spin chirality fluctuation of the Nd spins in addition to the ordered component. 

In this paper we study $\sigma_{xy}$ and also the longitudinal conductivity $\sigma_{xx}$ in the presence of the fluctuating spin chirality of the localized spins. The purpose of this study is twofold. One is to analyze the experiments in Nd${}_2$Mo${}_2$O${}_7$ semiquantitatively. The other is to reveal the condition for the spin chirality fluctuation to contribute to $\sigma_{xy}$ in terms of the lattice structure and the anisotropy of the localized spins.

\begin{figure}[tbh]
\begin{center}\leavevmode
\epsfxsize=2.5cm
$$\epsffile{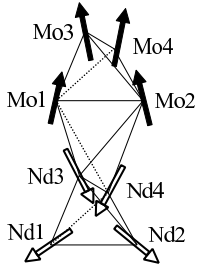}$$
\end{center}
\caption{Low-temperature spin structure of the Mo itinerant spins and the Nd localized spins in a unit cell for Nd${}_2$Mo${}_2$O${}_7$. }
\label{fig:tetrahedrons}
\end{figure}

To be explicit, we start with a model for Nd${}_2$Mo${}_2$O${}_7$. This compound has a pyrochlore structure with two sublattices of the tetrahedrons of Nd and Mo displaced by half a lattice constant. The Nd atoms yield $J=5/2$ localized spins of $4f$ electrons, while the Mo atoms two itinerant $d$ electrons of the $t_{2g}$ orbitals per site. The localized spins are well described by a spin-ice model with uniaxial spin anisotropy specified by either parallel or antiparallel direction from vertex to center of mass of each tetrahedron~\cite{Fukazawa02}. We treat these localized spins as the spin-$1/2$ Ising spins. The nearest-neighbor pairs of $d$ electrons and localized spins are coupled by an antiferromagnetic exchange interaction $J_{fd}$~\cite{J_fd}. Then, our Hamiltonian reads
\begin{eqnarray}
{\cal H}&=&-t\sum_{\langle {\bf x}_i,{\bf x}_{i'}\rangle,\alpha}c^\dagger_{{\bf x}_i,\alpha}c_{{\bf x}_{i'},\alpha}-\sum_{{\bf x}_i}(\mu n_{{\bf x}_i}+h_{\rm eff}s^z_{{\bf x}_i})
\nonumber\\
&&{}+J_{fd}\sum_{\langle {\bf x}_i,{\bf X}_j\rangle}\vec{s}_{{\bf x}_i}\cdot\vec{n}_jS^{z_j}_{{\bf X}_j}.
\end{eqnarray}
$c_{{\bf x}_i,\alpha}$, $c^\dagger_{{\bf x}_i,\alpha}$, $n_{{\bf x}_i}$ and 
$\vec{s}_{{\bf x}_i}\equiv \frac{1}{2}\sum_{\alpha,\alpha'}c^\dagger_{{\bf x}_i,\alpha} \vec{\sigma}_{\alpha\alpha'}c_{{\bf x}_i,\alpha'}$ are the annihilation, creation, number and spin operators of conduction electrons on a Mo site ${\bf x}_i$ with a crystallographic index $i$ in the Mo tetrahedron shown in Fig.~\ref{fig:tetrahedrons}, respectively. $\alpha$ is a spin index. $\vec{\sigma}$ is the vector form of the Pauli matrices. $\vec{n}_j$ is the unit vector pointing to the center of mass of the tetrahedron for the Nd spin with a crystallographic index $j$. $S^{z_j}_{{\bf X}_j}=\vec{S}_{{\bf X}_j}\cdot\vec{n}_j$ with the spin-$1/2$ localized spin operator $\vec{S}_{{\bf X}_j}$ at a Nd site ${\bf X}_j$ with an index $j$. $\sum_{\langle\cdots\rangle}$ denotes the summation over the nearest neighbors. The chemical potential $\mu$ is self-consistently determined with $\langle n_i\rangle$. The itinerant ferromagnetism occurs due to a double-exchange mechanism in the degenerate Mo 4$d$ orbitals. For simplicity we consider the single band model and treat the ferromagnetism in terms of the effective magnetic field $h_{\rm eff}$ in the $z$ direction only for the conduction electrons~\cite{h_eff}. Besides, we assume a reduced uniform electron transfer integral $t$ between the nearest-neighbor $d$ electrons. In the double exchange mechanism~\cite{AndersonHasegawa55} ignored here, the coherence temperature of conduction electrons is given by $T_{\rm C}$. We take into account this effect by choosing a reduced value $t=100$ K, which is an order of magnitude smaller than the bare transfer integral and is similar to $T_{\rm C}\sim90$ K~\cite{t_eff}.

We write the Dyson equation for the electron Green's function in the 
$8\times8$-matrix form as 
$\mib{G}^{-1}(\omega+i\delta,{\bf k})=\mib{G}_{\rm MF}
(\omega+i\delta,{\bf k})-\mib{\Sigma}(\omega+i\delta,{\bf k})$ 
with a positive infinitesimal number $\eta$, where
\begin{eqnarray}
&&G^{-1}_{{\rm MF} i,i'|\alpha,\alpha'}(\omega+i\eta,{\bf k})
\nonumber\\
&&=\left[(\omega+\mu+\sigma^z_{\alpha,\alpha'}h_{\rm eff}/2+i\eta)
\delta_{i,i'}+2tv_{{\bf k},i,i'}\right]\delta_{\alpha,\alpha'}
\nonumber\\
&&{}-J_{fd}\delta_{i,i'}\sum_jv_{{\bf q}=0,i,j}\langle S^{z_j}
\rangle\vec{n}_j\cdot\vec{\sigma}_{\alpha,\alpha'}.
\label{eq:G_MF}
\\
&&\Sigma_{i,i'|\alpha,\alpha'}(\omega+i\eta,{\bf k})
=J_{fd}^2\frac{T}{N}\sum_{{\bf q},j,j',
\bar{\alpha},\bar{\alpha}'}\!\!\!v_{{\bf q},i,j}
\chi_{j,j'}({\bf q})v_{-{\bf q},i',j'}
\nonumber\\
&&{}\times \left(\vec{\sigma}_{\alpha,\bar{\alpha}}\cdot\vec{n}_j\right)G_{i,i'|\bar{\alpha},\bar{\alpha}'}(\omega+i\eta,{\bf k}-{\bf q})\left(\vec{\sigma}_{\bar{\alpha}',\alpha'}\cdot\vec{n}_{j'}\right)
\label{eq:Sigma}
\end{eqnarray}
$v_{{\bf q},i,j}$ is the dimensionless form factor that describes the nearest-neighbor couplings between the Mo and Nd sites. $\chi_{i,j}({\bf q})$ is the wave-vector dependent susceptibility of the localized spins defined by $\chi_{j,j'}({\bf q})=(N/T)\langle \delta S^{z_j}_{{\bf q},j}\delta S^{z_{j'}}_{-{\bf q},j'}\rangle$ with the temperature $T$, $\delta S^{z_j}_{{\bf q},j}\equiv S^{z_j}_{{\bf q},j}-\langle S^{z_j}_j\rangle\delta_{{\bf q},0}$ and $S^{z_j}_{{\bf q},j}=\frac{1}{N}\sum_{{\bf X}_j}S^{z_j}_{{\bf X}_j}e^{-i{\bf q}\cdot{\bf X}_j}$. We note that these localized Ising spins are static. 

First, we neglect the correlations of the localized spins. Then, Eq. (\ref{eq:Sigma}) is exact with $\langle S^{z_j}_j\rangle=0$ and $\chi_{j,j'}({\bf q})=\delta_{j,j'}/4T$. Defining the spin components of the Green's function and the self-energy part as $\vec{\cal G}_{i,i'}=\sum_{\alpha,\alpha'}\vec{\sigma}_{\alpha',\alpha}G_{i,i'|\alpha,\alpha'}$ and $\vec\Sigma_{i,i'}=\sum_{\alpha,\alpha'}\vec{\sigma}_{\alpha',\alpha}\Sigma_{i,i'|\alpha,\alpha'}$, respectively, we obtain
\begin{eqnarray}
&&\vec{\Sigma}_{i,i'}(\omega+i\eta,{\bf k})=\frac{J_{fd}^2}{N}\sum_{{\bf k}',j}v_{{\bf k}-{\bf k}',i,j}v_{-{\bf k}+{\bf k}',i',j}
\nonumber\\
&&\times\!\left[2(\vec{\cal G}_{i,i'}(\omega+i\eta,{\bf k}')\cdot\vec{n}_j)\vec{n}_j-\vec{\cal G}_{i,i'}(\omega+i\eta,{\bf k}')\right].
\label{eq:vecSigma}
\end{eqnarray}
Under the itinerant ferromagnetism, ${\cal G}$ has a nonvanishing $z$ component. Since all of $\vec{n}_j$'s deviate from the $z$ direction to form a noncoplanar spin structure and every Mo site is linked to three pairs of Nd sites with a different index $j$, the first term in the square bracket in Eq. (\ref{eq:vecSigma}) yields a tilting of the conduction electron spins like the solid arrows in Fig~\ref{fig:tetrahedrons}. This tilting gives a noncoplanar spin structure of the conduction electrons and a spin chirality. 

These results can be interpreted as follows. One can consider three stages of the approximations. (a) The first is the simplest one, namely to take the average of the localized spins $\langle\vec{S}_j\rangle$ first, and to calculate the spin chirality for these averaged moments~\cite{YasuiSato}, which is zero in the disordered state. (b) The second is to consider the average of the spin chirality for the localized moments, i.e., $\langle Ch^{\rm loc.}_{ijk}\rangle = \langle \vec{S}_i \cdot \vec{S}_j \times \vec{S}_k\rangle$ which is again zero for the disordered case. (c) However, what is relevant to the transport properties is the spin chirality of the conduction electrons, namely that of Mo spins ${\vec s}_i$. One should then take into account the ferromagnetic moment along $z$ direction. Therefore only the $x$ and $y$ components are determined by the localized Nd spins. In this case one can show that $\langle Ch^{\rm cond.}_{ijk}\rangle = \langle \vec{s}_i \cdot \vec{s}_j \times \vec{s}_k\rangle$ does not vanish. 

Now we consider the condition for the applicability of this fluctuation induced $\sigma_{xy}$. First we replace the localized Ising spins by the classical Heisenberg spins. Then, we obtain
$
\vec{\Sigma}_{i,i'}(\omega+i\eta,{\bf k})=-\frac{J_{fd}^2}{N}\sum_{{\bf q},j}v_{{\bf q},i,j}v_{-{\bf q},i',j}\vec{\cal G}_{i,i'}(\omega+i\eta,{\bf k}-{\bf q})
$
in the local moment limit. Here, the tilting effects of the conduction electron spins is smeared out, leading to $\sigma_{xy}=0$. This is also consistent with the experiment in Gd${}_2$Mo${}_2$O${}_7$, where Gd${}^{3+}$ has no orbital angular moment and hence the Ising anisotropy is absent, showing no enhancement of $\sigma_{xy}$ towards low temperatures~\cite{tokura}. Next we consider another crystal structure, i.e., the cubic lattice. The Ising spins are put on each lattice position pointing toward the center of the cubes. Similar analysis gives again $\sigma_{xy}=0$ in this case. It is because the periodicity of the crystal excludes the uniform magnetic field produced by the gauge flux, and all the elementary plaquettes are equivalent.

We can extend these arguments to more general single-orbital lattice models under the itinerant ferromagnetism with the net magnetization along the $z$ direction, to obtain the necessary condition as: (I-i) When the translational symmetry is not broken, the unit cell of the Bravais lattice must contain no less than three localized spins with different spin anisotropies and three itinerant spins where three are located on different positions for both localized and itinerant spins. (I-ii) For each itinerant spin, the directly coupled localized spins should satisfy the following two conditions; (I-iia) the spin inversion symmetry with respect to the $xy$ plane is broken and (I-iib) there remains a finite inplane component of the summation of $(n^x_{j,a},n^y_{j,a})n^z_{j,a}$ over the spins $j$ and the internal degrees of freedom of each spin $a$. For example, in the case of the classical spin without anisotropy or with the easy plane perpendicular to the $z$ axis, $a$ corresponds to the degrees of freedom over (half of) the unit sphere or a circle, respectively. Then, the integral over this variable gives zero and $\sigma_{xy}=0$. (II) If the translational symmetry is broken and the unit cell is enlarged, then the new unit cell should contain the part satisfying (I-i) and (I-ii). The above conditions are also applicable to the case where the Ising variables are ordered.
\begin{figure}[tbh]
\begin{center}\leavevmode
\epsfxsize=5.5cm
$$\epsffile{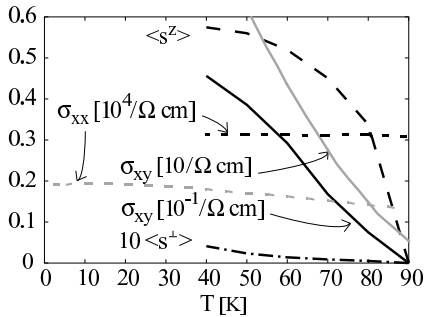}$$
\end{center}
\caption{Black lines represent numerical results in the local moment limit of the localized spins for $\langle s^z\rangle$, $\langle s^\perp\rangle$, $\sigma_{xx}$ and $\sigma_{xy}$ after rescaling {\protect\cite{transform}}. $J_{fd}=80$ K {\protect\cite{J_fd}} and $\langle n_i\rangle=0.7$ are adopted. Grey lines represent experimental results for $\sigma_{xy}$ and $\sigma_{xx}$ at an applied field along the $z$ direction $0.5$ T by Taguchi et al. {\protect\cite{TaguchiScience01}}.}
\label{fig:local}
\end{figure}

Therefore Nd${}_2$Mo${}_2$O${}_7$ is a special case where the chirality fluctuation of the localized spins contributes to the AHE. The fluctuation induced $\sigma_{xy}$ qualitatively explains the experimental results in Nd${}_2$Mo${}_2$O${}_7$ that $\sigma_{xy}$ increases below $T_{\rm C}$ even when a ordered moment and spin chirality of the Nd localized spins are negligibly small~\cite{TaguchiScience01}. We have obtained self-consistent solutions of the Green's function and calculated the longitudinal and transverse conductivities $\sigma_{xx}$ and $\sigma_{xy}$ by neglecting the vertex corrections. After rescaling the numerical results into the system with $\langle n_i\rangle=2$, and the $J=5/2$ localized spins~\cite{transform} for comparison with experiments, results for the ferromagnetic polarization of the conduction electrons $\langle s^z\rangle$ produced by $h_{\rm eff}$, amplitude of the inplane spin component of the conduction electrons $\langle s^\perp\rangle$, $\sigma_{xx}$ and $\sigma_{xy}$ are shown in Fig.~\ref{fig:local}. Here, with increasing $\langle s^z\rangle$ below $T_{\rm C}=90$ K, $\sigma_{xy}$ grows in the absence of correlations of the localized spins, although it is still two digits smaller than the experimental results~\cite{TaguchiScience01}. Since the quasiparticle damping rate $\gamma\propto J_{fd}^2/t$ has no prominent $T$ dependence in this local-moment case, $\sigma_{xx}$, which is proportional to $(t/J_{fd})^2$, is also temperature independent. Moreover, the magnitude of $\sigma_{xx}$ is similar to experimental results. This indicates that the present choice of $J_{fd}/t=0.8$ is almost consistent with experiments. 

\begin{figure}[tbh]
\begin{center}\leavevmode
\epsfxsize=5.0cm
$$\epsffile{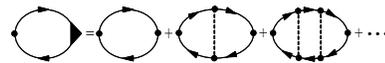}$$
\end{center}
\caption{Diagrammatic representation of $V_{j,j'}({\bf q})$. Solid lines, dashed lines and dots represent the electron Green's function $G_{i,i'|\alpha,\alpha'}({\bf k},\omega+i\eta)$, spin correlation function $T\chi_{j,j'}({\bf q})$, and the coupling $2J_{fd}v_{{\bf q},i,j}\vec{\sigma}\cdot\vec{n}_j\delta_{i,i'}$. The irreducible vertex corresponds to the dashed line with two dots at the ends.}
\label{fig:diagram}
\end{figure}

Next, to make the comparison with experiments in Nd${}_2$Mo${}_2$O${}_7$ more quantitatively, we include correlations of the localized spins, which enhance $\sigma_{xy}$. In the present case, the localized spins interact with each other through the conduction electrons, as in RKKY interaction. $\chi_{j,j'}({\bf q})$ is calculated by solving the following equations:
\begin{eqnarray}
\chi^{-1}_{j,j'}({\bf q})&=&4T\delta_{j,j'}\cosh^2\frac{\beta\varepsilon_j}{2}+2J_{fd}V_{j,j'}({\bf q}),
\label{eq:chi2}
\\
V_{j,j'}({\bf q})&=&\frac{1}{N}\sum_{i}v_{{\bf q},i,j}{\mit\Delta}\langle \vec{s}_{{\bf q},i}\rangle\cdot\vec{n}_j/{\mit\Delta}\langle S^{z_{j'}}_{{\bf q},j'}\rangle.
\label{eq:V}
\end{eqnarray}
with the ordered component of the localized spin determined by $\langle S^{z_j}_j\rangle=\frac{1}{2}\tanh(\varepsilon_j/2T)$ with $\varepsilon_j=2J_{fd}\sum_iv_{{\bf q}=0,i,j}\langle\vec{s}_{{\bf q}=0,i}\rangle\cdot\vec{n}_j$. The effective interaction $V_{j,j'}({\bf q})$ via the conduction electrons contains the function ${\mit\Delta}\langle \vec{s}_{{\bf q},i}\rangle$, which represents the linear response of the itinerant spin with a wave vector ${\bf q}$ and a crystallographic index $i$ to the ordered component of the localized spin with a wave vector ${\bf q}$ and an index $j'$. Then, calculation of $V_{j,j'}({\bf q})$ corresponds to the sum of the ladder diagrams shown in Fig.~\ref{fig:diagram}. This term enhances short-ranged correlations of the localized spins. The short-ranged correlations reflect fluctuations around the uniform two-in, two-out ordered state. We approximately calculate $V_{j,j'}({\bf q})$ by replacing the momentum-dependent irreducible vertex with its summation over the momentum. Namely, mode-mode couplings that mix collective excitations with different momenta are neglected. Then, a newly obtained $\chi_{j,j'}({\bf q})$ is substituted into Eq. (\ref{eq:Sigma}) to calculate a new Green's function until convergence is reached.

\begin{figure}[tbh]
\begin{center}\leavevmode
\epsfxsize=5.5cm
$$\epsffile{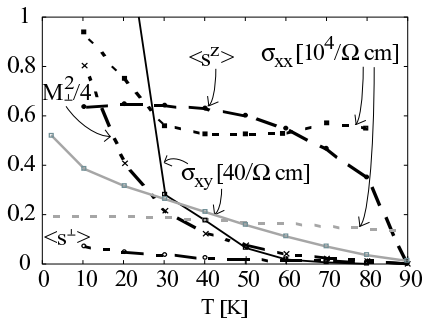}$$
\end{center}
\caption{Black lines represent numerical results after rescaling {\protect\cite{transform}} where correlations of the localized spins are taken account. $J_{fd}=40$ K {\protect\cite{J_fd}} and $\langle n_i\rangle=2/3$ are adopted. Grey lines represent experimental results for $\sigma_{xy}$ and $\sigma_{xx}$ at an applied field along the $z$ direction $0.5$ T by Taguchi et al. {\protect\cite{TaguchiScience01}}.}
\label{fig:RPAr0}
\end{figure}

Numerical results after the rescaling procedure~\cite{transform} are shown in Fig.~\ref{fig:RPAr0}. As $\langle s^z\rangle$ increases below $T_{\rm C}$, the transverse component of the magnetization $M_\perp$, $\langle s^\perp\rangle$ and $\sigma_{xy}$ increase. At intermediate temperatures around 50 K, an enhancement from the results without short-ranged correlations is more than 10. There is a crossover around $T_*\sim J_{fd}$ that characterizes a shift from dominant short-ranged components of the localized spin correlations at higher temperatures to long-ranged components at lower temperatures. At lower temperatures, $\sigma_{xy}$ exhibits a larger enhancement compared with experiments~\cite{TaguchiScience01}. These results indicate that inclusion of correlations, not only the long-ranged order but also short-ranged correlation of the localized spins enhance $\sigma_{xy}$. Although the present choice of $J_{fd}/t=0.4$ is half the previous one in Fig.~\ref{fig:local}, $\sigma_{xx}$ remains of the same order of magnitude as the experimental results~\cite{TaguchiScience01}. In a simple single-band case coupled to classical spins, $\gamma$ increases logarithmically with increasing characteristic length of spin fluctuations $\xi$. In the present case, molecular fields produced by the itinerant spins prevent $\xi$ from diverging. Then, below $T_*$, $\xi$ turns to decrease. This yields a shallow minimum of $\sigma_{xx}$ around $T_*$. The obtained $T_*$ inferred from the rapid increase of $M_\perp^2$ is similar to the experimental result~\cite{TaguchiScience01}. Since $T_*$ is mainly determined from $J_{fd}$, $J_{fd}\sim40$ K is also consistent with experiments.

Lastly, we mention relations of the present study to other recent theoretical ones. We have shown, in contrast to Ref.~\cite{YeKimMillisShraimanMajumdarTesanovic99}, that there is no uniform gauge flux, i.e., $\int_{\rm unit cell}\langle \vec{b}\rangle\, d{\bf x}=\vec{0}$. Hence the microscopic lattice structure is crucial. From this viewpoint, the single-band model on a cubic lattice does not show finite $\sigma_{xy}$ even though the second-neighbor hopping is introduced in contrast to Ref.~\cite{Lyanda-Geller01}. Tatara and Kawamura obtained $\sigma_{xy}$ proportional to the uniform spin chirality of the localized spins within an adiabatic approximation in the background of the static localized spins~\cite{TataraKawamura02}. Their approximation corresponds to the second stage of approximations (b) mentioned above. Their theory works in the dirty case $J\tau\ll1$ as in spin-glass systems with an exchange interaction $J$ and electron relaxation time $\tau$, while the present theory works even for $J_{fd}/t\leq 1$ as in Nd${}_2$Mo${}_2$O${}_7$~\cite{TaguchiScience01}. 

In summary, we have studied a spin chirality mechanism of the AHE. We have found that in itinerant ferromagnets with localized spins forming superstructures with different uniaxial spin anisotropies, the spin chirality mechanism works even without the correlations of the localized spins. Both long-ranged and short-ranged correlations of the localized spins enhance the Hall conductivity. Full quantitative explanations of the AHE in Nd${}_2$Mo${}_2$O${}_7$ require including the $t_{2{\rm G}}$ orbitals and Coulomb interactions, which is left for future studies.

The authors are grateful to Y. Tokura, M. Sato, Y. Maeno and G. Tatara for 
discussion and comments. We would like to thank Y. Taguchi for providing us 
experimental data.

\end{multicols}


\begin{references}
\bibitem{h_eff}
To mimic the exchange field arising from the degenerate electrons, we assume $h_{\rm eff}=h_0\tilde{s}^z$ with $h_0=1000$ K and determine $\tilde{s}^z$ from the solution to $\tilde{s}^z=\frac{1}{2}\tanh(2T_{\rm C}\tilde{s}^z/T)$ and $T_{\rm C}=90$ K. This roughly reproduces the $T$ dependence of the ferromagnetic magnetization.
\bibitem{t_eff}
The effective reduced transfer integral $t=100$ K is also similar to the value estimated from the analysis on the Drude weight in the optical conductivity in Sm${}_2$Mo${}_2$O${}_7$~\cite{TaguchiTokura_unpublished} ($T_{\rm C}=73$ K~\cite{TaguchiTokura99}), i.e., $t_{\rm exp}\sim 0.025 \hbar^2/mb^2\sim75$ K with the free-electron mass $m$ and the Mo-Mo length $b=3.7\times10^{-8}$ cm.
\bibitem{transform}
For comparison with experiments, numerical reuslts in the present model with the number of electrons per site $\langle n_i\rangle$ and the spin-$1/2$ localized spins are plotted after the operations; $\langle s^z\rangle=\langle s^z_i\rangle(2/\langle n_i\rangle)$, $\langle s^\perp\rangle=\langle s^\perp_i\rangle(2/\langle n_i\rangle)$, $M_\perp=5\langle S^\perp_i\rangle-\langle s^\perp_i\rangle(2/\langle n_i\rangle)$, $\sigma_{xx}\to\sigma_{xx}(2/\langle n_i\rangle)$ and $\sigma_{xy}\to\sigma_{xy}(2/\langle n_i\rangle)$ where $\langle S^\perp_j\rangle$ is the amplitude of the global $xy$-plane component of $\langle\vec{S}^{z_j}_j\rangle$.
\end{references}
\end{document}